\documentclass[a4paper,alpha-refs]{aejstyles}

\journal{aej}

\usepackage{graphicx}
\usepackage{siunitx}
\usepackage{subcaption}

\usepackage{booktabs}
\usepackage{amssymb}

\usepackage[left]{lineno}

\usepackage{flushend}

\title{Bringing Lecture-Tutorials Online: An Analysis of A New Strategy to Teach Planet Formation in the Undergraduate Classroom}

\author[1,2]{Haylee N.\ Archer}
\author[1]{Molly N.\ Simon}
\author[1]{Chris Mead}
\author[3]{Edward E. Prather}
\author[1]{Mia Brunkhorst}
\author[1]{Diana Hunsley}

\affil[1]{School of Earth and Space Exploration, Arizona State University, Tempe, AZ, USA}
\affil[2]{Lowell Observatory, Flagstaff, AZ, USA}
\affil[3]{Department of Astronomy, University of Arizona, Tucson, AZ, USA}

\papercat{Research Article}

\runningauthor{Archer et al.}

\jvolume{04}
\jnumber{1}
\jyear{2024}

\begin{document}

\begin{frontmatter}
\maketitle
\begin{abstract}
Previous studies conclusively show that pencil-and-paper lecture-tutorials (LTs) are incredibly effective at increasing student engagement and learning gains on a variety of topics when compared to traditional lecture. LTs in astronomy are post-lecture activities developed with the intention of helping students engage with conceptual and reasoning difficulties around a specific topic with the end goal of them developing a more expert-like understanding of astrophysical concepts. To date, all astronomy LTs have been developed for undergraduate courses taught in-person. Increases in online course enrollments and the COVID-19 pandemic further highlighted the need for additional interactive, research-based, curricular materials designed for online classrooms. To this end, we developed and assessed the efficacy of an innovative, interactive LT designed to teach planet formation in asynchronous, online, introductory astronomy courses for undergraduates. We utilized the Planet Formation Concept Inventory to compare learning outcomes between courses that implemented the new online, interactive LT, and those that used either a lecture-only approach or utilized a standard pencil-and-paper LT on the same topic. Overall, learning gains from the standard pencil-and-paper LT were statistically indistinguishable from the in-person implementation of the online LT and both of these conditions outperformed the lecture-only condition. However, when implemented asynchronously, learning gains from the online LT were lower and not significantly above the lecture-only condition. While improvements can be made to improve the online LT in the future, the current discipline ideas still outperform traditional lecture, and can be used as a tool to teach planet formation effectively. 

\end{abstract}


\end{frontmatter}


\section{Introduction}\label{sec:intro}

The United States requires all 4-year college students to complete at least one semester-long science course, enrolling hundreds of thousands of students in general education science courses every year \citep{rudolph_national_2010}. For non-science majors (students who do not intend to pursue an undergraduate science degree), these courses are often their last formal science instruction, which influences their personal viewpoints and civil engagement with scientific issues \citep{hobson_surprising_2008}. In a world where the internet and other media offer conflicting information on scientific research, the importance of scientific and data literacy is at an all-time high. Developing classroom materials for these courses that address common preconceptions and increase student understanding is essential for creating ``competent outsiders'', non-scientists who understand how science relates to local or personal issues without relying on specific scientific concepts learned in the classroom \citep{feinstein_outside_2013}. An introductory, semester-long, astronomy course for non-science majors, commonly referred to as ASTRO 101 is often taken as this general science elective. As such, it is especially important to ensure that students leave ASTRO 101 with a better understanding of our place in the Universe before becoming active members of society who will engage with broader scientific concepts outside of the classroom.

To date, traditional lecture is the dominant form of undergraduate instruction, but several cross-disciplinary studies have shown that the implementation of active learning strategies leads to higher student learning outcomes \citep{chi_icap_2014,freeman_active_2014}. While the concept of active learning is broadly defined \citep{lombardi_curious_2021}, we define active learning to mean requiring students to interact with and think deeply about classroom material in a meaningful way, as opposed to traditional lecture where students passively receive information. One well-researched active learning strategy in ASTRO 101 is the Lecture-Tutorial. Lecture-Tutorials (LTs) are worksheets designed to supplement lecture, and typically require that students work in small, collaborative groups. The LTs consist of a series of questions that build on one another, and address common conceptual and reasoning difficulties that arise as students learn about a variety of topics in astronomy. In the domain of ASTRO 101 courses, LTs have been used for decades, resulting in significant increases in student learning on a variety of topics \citep[e.g.][]{prather_research_2004,LoPresto_2009,wallace_study_2012,lombardi_curious_2021}.

Far less research has been conducted on the use of active learning strategies like LTs in online astronomy courses, and there is a scarcity of learner-centered, research-based instructional materials designed for the online student population \citep{simon_new_2022}. This insufficiency was further highlighted when the COVID-19 pandemic required courses traditionally taught in-person to pivot online with little notice. Even prior to the COVID-19 pandemic, online course enrollments have increased exponentially due to online courses’ accessibility and appeal \citep[e.g.][]{allen_changing_2013,cooper_diagnosing_2019}. Students benefit from the ability to enroll in courses without the need to commute to a physical classroom, expanding access to higher education to students who may otherwise have difficulty attending courses in-person (e.g.\ caretakers, active military personnel, and full-time employees). Increases in online ASTRO 101 enrollments coupled with limited active learning-based curricular materials accessible in the online format motivated the development of an online LT, and a research effort to assess whether an online LT will lead to student learning that is consistent with what has been seen with the pencil-and-paper LTs. 

We created an online LT specifically for ASTRO 101 courses that was designed to actively engage students in learning about the topic of planet formation. We modeled the online LT after the Planet Formation Lecture-Tutorial (PFLT), a version of which is published in \textit{Lecture-Tutorials for Introductory Astronomy, 4th Edition} \citep{prather_lecture_2021}. The discovery and characterization of over 5,000 planets outside of our Solar System (exoplanets) highlights the importance of integrating planet formation into the ASTRO 101 curriculum. By learning about how planets and planetary systems form, students gain a better understanding of the origin and evolution of both our own Solar System and the discovered planetary systems beyond. Exoplanet discovery and characterization is one of the most active areas of research in astronomy, and it is important that ASTRO 101 students have a preliminary understanding of planet formation in order to make comparisons between exoplanetary systems and our own planetary neighborhood.

Through this online LT development work we explored the following questions: 1. Is the pedagogical approach employed for pencil-and-paper LTs enhanced when converted to a digital version (which includes additional interactive elements)? 2. How do student learning gains compare between the new online LT and the traditional LT, especially when considering the extent of student learning in online asynchronous courses? This paper is organized as follows: we present an overview of the PFLT in Section \ref{sec:pflt} and the translation of the PFLT to our Planet Formation Online Lecture-Tutorial (PFOLT) in Section \ref{sec:pfolt}. In Section \ref{sec:methods} we introduce our study participants and describe the assessment used in the study, along with our analysis methods. We present our results and discuss our findings in Section \ref{sec:results} and Section \ref{sec:discuss} respectively. Finally, our conclusions and opportunities for future work are presented in Section \ref{sec:summary}.

\section{Overview of the Planet Formation Lecture-Tutorial (PFLT)}\label{sec:pflt}

The format and question sequence of the PFLT was modeled after the process used to develop LTs on other disciplinary topics \citep[e.g.][]{prather_research_2004,wallace_using_2016,wallace_new_2021}. 

The activity employs a variety of representations (graphs, data tables, drawings, etc.) paired with carefully sequenced, questions and tasks intended to engage students in disciplinary discernment and increase their fluency with the topic of planet formation \citep{french_systematic_2020,simon_new_2022}. 
The PFLT is intended to be administered as a 25-30 minute, small-group (2-3 student per group) activity following a lecture on the topic of planet formation and relevant sub-topics (e.g.\ gravity, angular momentum, and condensation of the elements). After completing the PFLT, students should be able to:

\begin{itemize}
    \item Distinguish the formation of out Solar System from the formation of the Universe
    \vspace{0.5em}
    \item Apply the relationship between distance from the Sun and condensation temperature to predict the composition of planets at a variety of locations
    \vspace{0.5em}
    \item Identify the location of the frost/snow line and its relationship to planetary composition
    \vspace{0.5em}
    \item Explain how it could be possible for a gas/ice giant planet to be found inside the frost/snow line of a hypothetical exoplanetary system
\end{itemize}

These learning outcomes and the overall content presented in the PFLT were informed by prior work investigating ASTRO 101 students’ conceptual and reasoning difficulties on the topic of planet formation \citep{simon_survey_2018}. Most notably was ASTRO 101 students’ inability to distinguish the formation of the Solar System from the formation of the Universe, despite the events being separated by more than nine billion years. To this end, the PFLT begins with a question sequence that culminates with a hypothetical student debate aimed to challenge students’ understanding of cosmological time. Hypothetical student debates are prevalent amongst LTs and model conversations free of science jargon between 2-3 students where one student presents a common reasoning difficulty and the other student challenges this reasoning difficulty in favor of a more scientifically accurate explanatory model. An example from the PFLT is below:  

\begin{quote}
    Student 1: I think the formation of the Universe and the formation of the Solar System are totally different events. The Universe formed billions of years before our Solar System.
    
    \vspace{1em}
    \noindent Student 2: I don’t think so. All of the material in the Universe was created during the Big Bang, so our Solar System must have formed when the Universe did, nearly 14 billion years ago.
    
    \vspace{1em}
    \noindent Do you agree or disagree with either or both of the students? Explain your reasoning.
\end{quote}

\begin{figure}[b!]  
\centering
\includegraphics[width=\columnwidth]{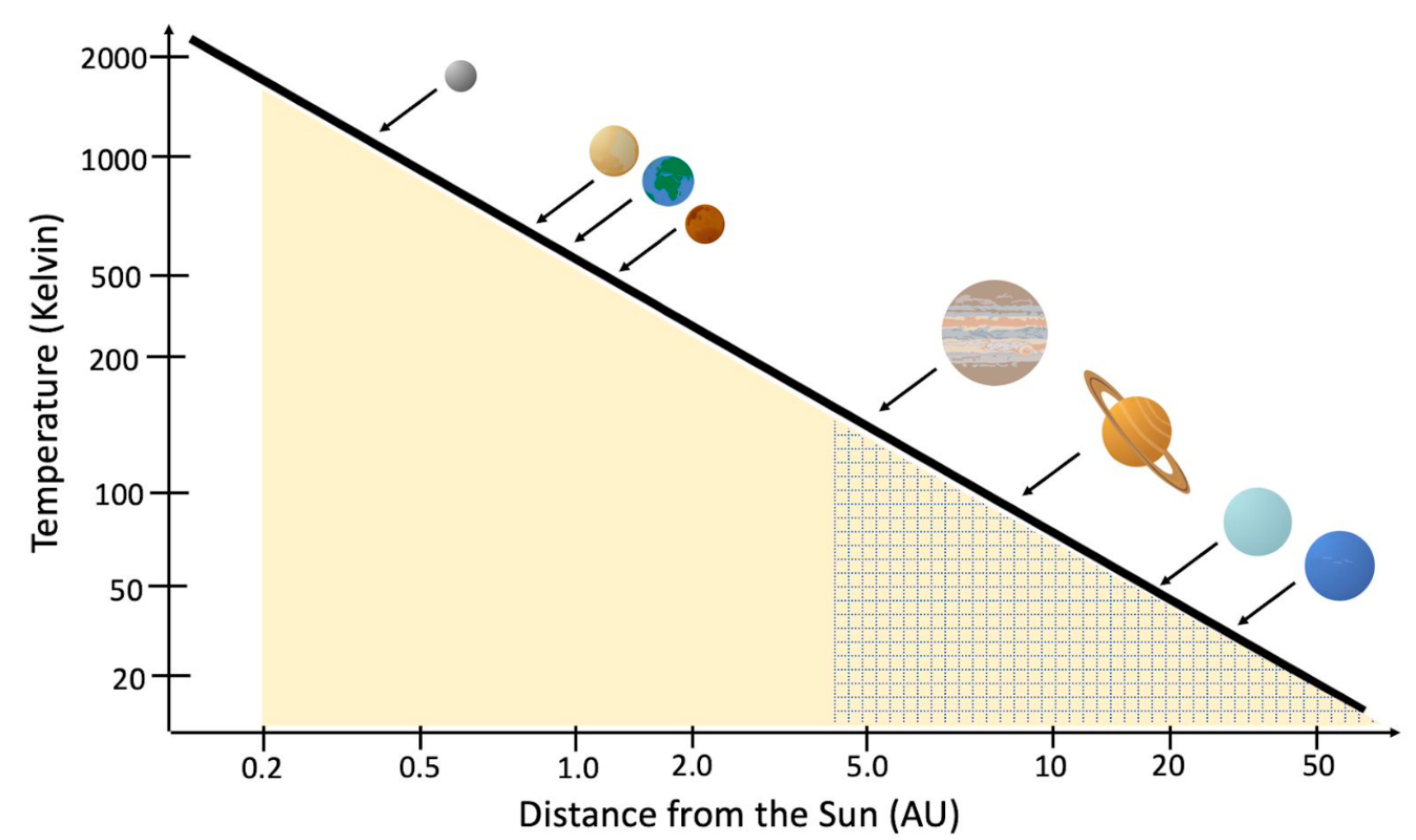}
\caption{Temperature in the protoplanetary disk at the time of planet formation versus distance from the Sun for our Solar System. The region (and temperature range) where rock and metals condense is shaded in pale yellow, and where hydrogen compounds condense to form ice is overlaid by a blue grid pattern. The relative locations of the planets are indicated by arrows. 
}\label{fig:pflt}
\end{figure}
In this particular example, learners are presented with an opportunity to challenge Student 2, who conflates the formation of the Universe with the formation of our Solar System. Requiring students to confront their own conceptual and reasoning difficulties head-on is an exceptionally valuable tool in promoting a metacognitive approach to learning (in which students cultivate an awareness of their thinking processes and how they learn) leading to more persisting conceptual change \citep{posner_accommodation_1982,prather_research_2004}.

Students then learn about the timeline of Solar System formation from cloud collapse to the formation of the Sun, the protoplanetary disk, planetesimals, and ultimately, planets. After a short question sequence highlighting the role of gravity in planet formation, the PFLT introduces the concept of condensation temperature, which is a focal point of the remainder of the LT. The condensation temperature component of the PFLT begins with a table consisting of the condensation temperature and relative abundances (mass \%) of hydrogen and helium gas, silicates (hereafter referred to as rock) and metals, and hydrogen compounds (e.g.\ water, methane, and ammonia) in the protoplanetary disk. Students are then shown a graph of the relationship between temperature in the disk at the time of planet formation (y-axis) and distance from the Sun (x-axis) for the planets in our Solar System (Figure \ref{fig:pflt}). The variables represented on the x and y-axes can be approximately represented with a power law for the early solar system. Although the relationship between temperature and distance in actuality is more complex, it is important that introductory students are able to understand at the most fundamental level that temperature in the disk decreases with distance from the central star. Purposefully displaying data in an accessible way is common for pedagogical discipline representations or PDRs. PDRs ``depict stylized physical scenarios and highlight discipline relationships that, while invaluable pedagogically, have little to no value to experts and professionals working in that field'' \citep[p. 2]{french_systematic_2020}. PDRs are often included in LTs due to their ability to assist students in developing stronger representational competence surrounding a given topic \citep{french_systematic_2020,volkwyn_developing_2020,simon_new_2022}. 

Students use Figure \ref{fig:pflt} (and the condensation temperature values presented in a corresponding table) to determine the range of distances in the protoplanetary disk over which rock and metals and hydrogen compounds condense during our Solar System’s formation. Students are then required to input the solid materials present at the location of each of the planets into a table where a column containing this information is intentionally left blank. 

\begin{figure}[t!]  
\centering
\includegraphics[width=\columnwidth]{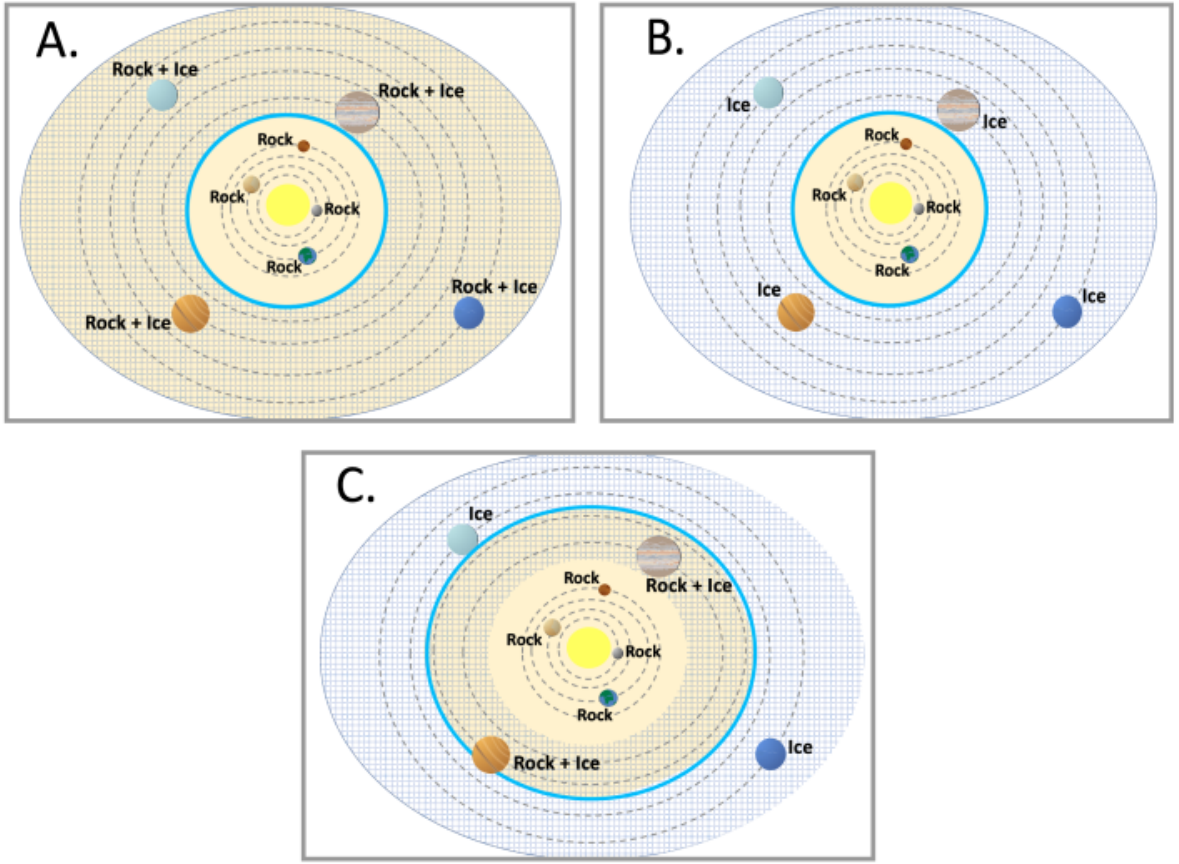}
\caption{Students are required to select which of the three diagrams (A, B, or C) most accurately represents the solid materials available to each of the planets in our Solar System during formation. Pale yellow corresponds to the region where solid rocks and metals condense, and the blue grid pattern is used to identify the region where hydrogen compounds condense into ice. A thick blue line (known as the snow line or frost line) is drawn to represent the location in the Solar System nearest the Sun where ices can begin to form.  
}\label{fig:frost}
\end{figure}

At this point in the PFLT, students are presented with a choice of three diagrams, one of which most accurately represents the distribution of solid material in our Solar System at the time of planet formation (Figure \ref{fig:frost}).

Through analyzing Figure \ref{fig:frost}, students integrate the information from multiple data representations (graph and table) to demonstrate an understanding that rocks and metals are able to condense throughout the protoplanetary disk, whereas hydrogen compounds condense only in the outer Solar System beyond the frost line. Next, students engage with a student debate intended to address any reasoning difficulties learners may still have with the relationship between condensation temperature, the frost line, and planetary composition. The student debate is structured as follows:

\begin{quote}
    Student 1: I think drawing ``C'' is correct because we know the Terrestrial planets are made of rock and Neptune and Uranus are ice giants so they will be the only planets made of just ice.

    \vspace{0.8em}
    \noindent Student 2: I agree with you, but I think you need to include Jupiter and Saturn as having some ice too, and based on the graph the blue frost line should be drawn closer to the Sun than Jupiter, so I think it’s drawing ``B''. 

    \vspace{0.8em}
    \noindent Student 3: I think you’re right that the frost line should be drawn closer to the Sun, but I think drawing ``A'' is correct because there were rocks and metals throughout the early Solar System but ice only formed past the frost line where we find the gas giant planets.

    \vspace{0.8em}
    \noindent Which student do you agree with? Which do you disagree with? Explain your reasoning.
\end{quote}

Through their peer discussions of the range of ideas presented in this student debate, learners have an opportunity to address the most prevalent reasoning difficulties on the topic of condensation temperature, namely that the frost line acts as a barrier between solid rocks/metals and ices, and that rocks and metals are only able to condense at distances inward of the frost line \citep{simon_survey_2018}. Next, students complete a short fill-in-the-blank section that helps learners develop the relationship between the availability of solids for a particular planet location and the differences in mass between the inner and outer planets in our Solar System. 

\begin{table}[t!]
\caption{Planetary parameters for three planets in a hypothetical exoplanetary system, sorted by planet mass.}\label{tab:planets}
\begin{tabularx}{\columnwidth}{l c c c}
\toprule
Planet & Mass & Distance from star & Atmosphere \\
Name & (Earth = 1) & (AU) & (Large/Small)\\
\midrule
A & 0.643 & 1.774 & Small \\
B & 11.34 & 6.482 & Large \\
C & 12.01 & 0.031 & Large \\
\bottomrule
\end{tabularx}
\end{table}
The PFLT concludes with a section that requires students to apply their knowledge of the relationship between condensation temperature, planet mass, and distance from the central star to a hypothetical exoplanetary system with three planets. Relevant properties for each of the planets in this system are listed in Table \ref{tab:planets}. 

The goal of this part of the activity is to identify which (if any) of the planets in the hypothetical exoplanetary system are at locations (relative to their host star) we would expect based on what they have learned about planet formation in our Solar System. Once students identify that Planet C is much closer to its host star than what would be expected of a giant planet, we introduce the final student debate of the PFLT. 

\begin{quote}
    Student 1: I think that the physics that explains where rock, ice, and gas would exist during the planet formation process doesn’t apply when we’re dealing with planets in other solar systems.

    \vspace{0.8em}
    \noindent Student 2: I disagree. I think that the locations where we’d expect to find rock, ice, and gas would be pretty much the same in every solar system. What I think happens is that the star gets way more massive after the solar system forms, and this pulls planets closer in towards the star.

    \vspace{0.8em}
    \noindent Student 3: We learned that essentially all the mass of a solar system is in the star already, and if it did get more massive it would pull all the planets inward, not just this one gas giant. I think these planets must be interacting with other objects in the solar system and that eventually causes the planet to move out of the position where it was originally forming.

    \vspace{0.8em}
    \noindent Which student do you agree with? Which do you disagree with? Explain your reasoning.
\end{quote}

This final debate introduces the concept that planets may move from the original locations in which they formed. Instructors are encouraged to use this final debate as a launching point to discuss Hot Jupiters, a class of giant exoplanet discovered at distances typically less than 0.1 AU from their host star. 
\begin{figure}[tb!]
\captionsetup[subfigure]{labelfont=bf}
\captionsetup[subfigure]{justification=centering}
\centering

\begin{subfigure}[b]{\columnwidth}
   \includegraphics[width=1\linewidth]{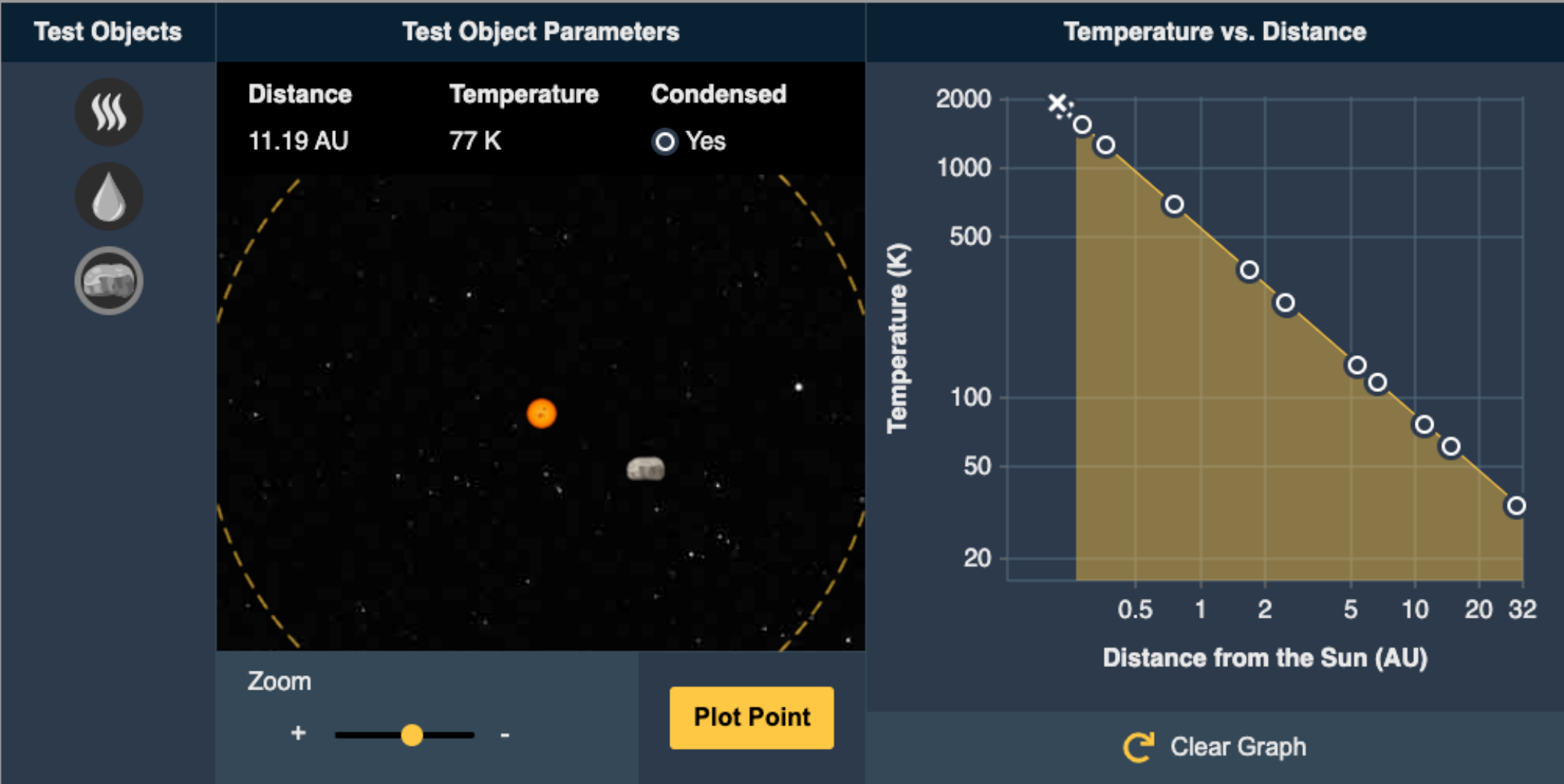}\vspace{-0.5em}
   \caption{}
   \vspace{1em}
   \label{fig:rock}
\vspace{-0.5em}
\end{subfigure}
\vspace{-0.5em}
\begin{subfigure}[b]{\columnwidth}
   \includegraphics[width=1\linewidth]{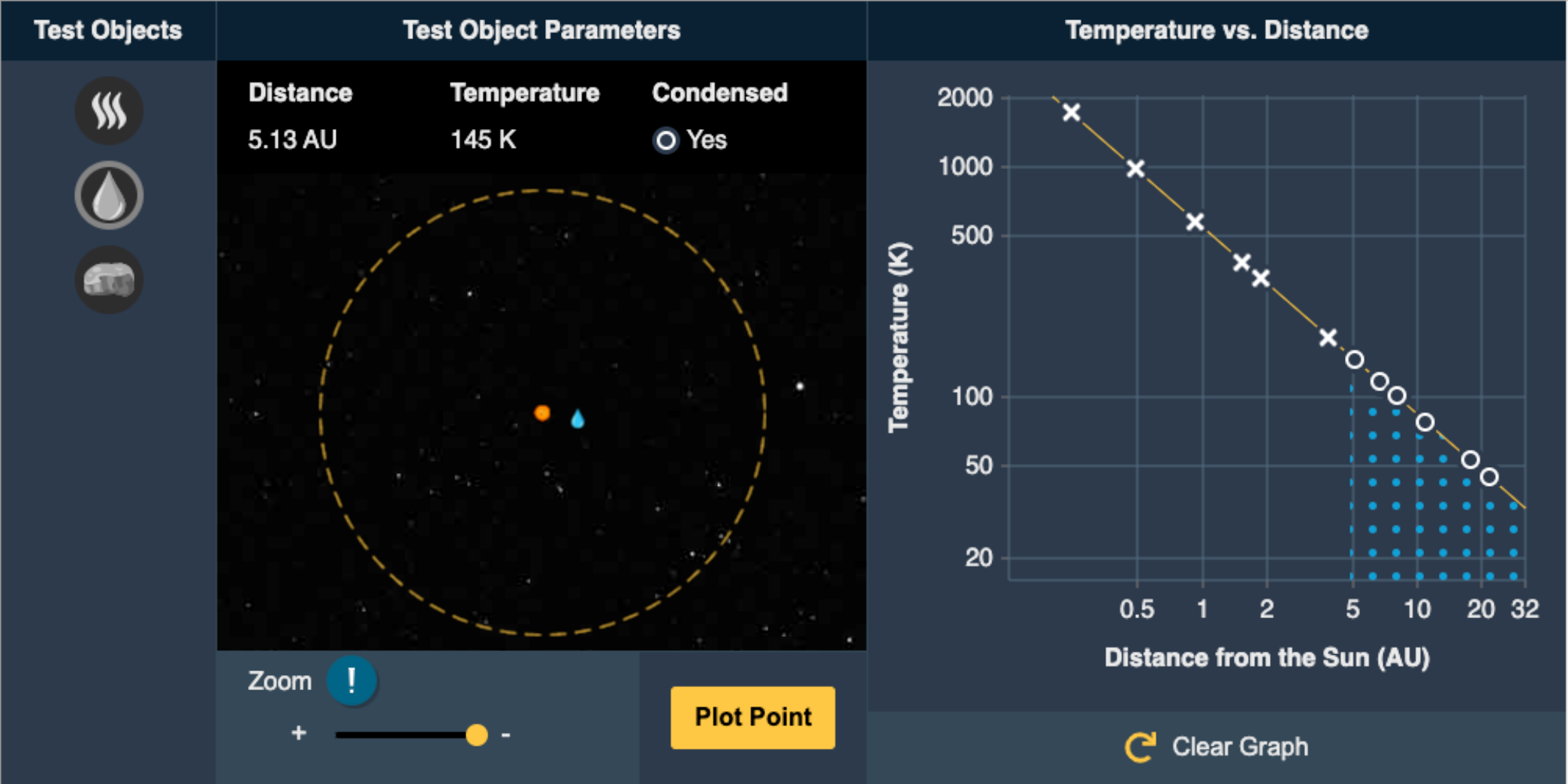}\vspace{-0.5em}
   \caption{}
   \vspace{1em}
   \label{fig:water}
\end{subfigure}
\vspace{-1em}
\begin{subfigure}[b]{\columnwidth}
   \includegraphics[width=1\linewidth]{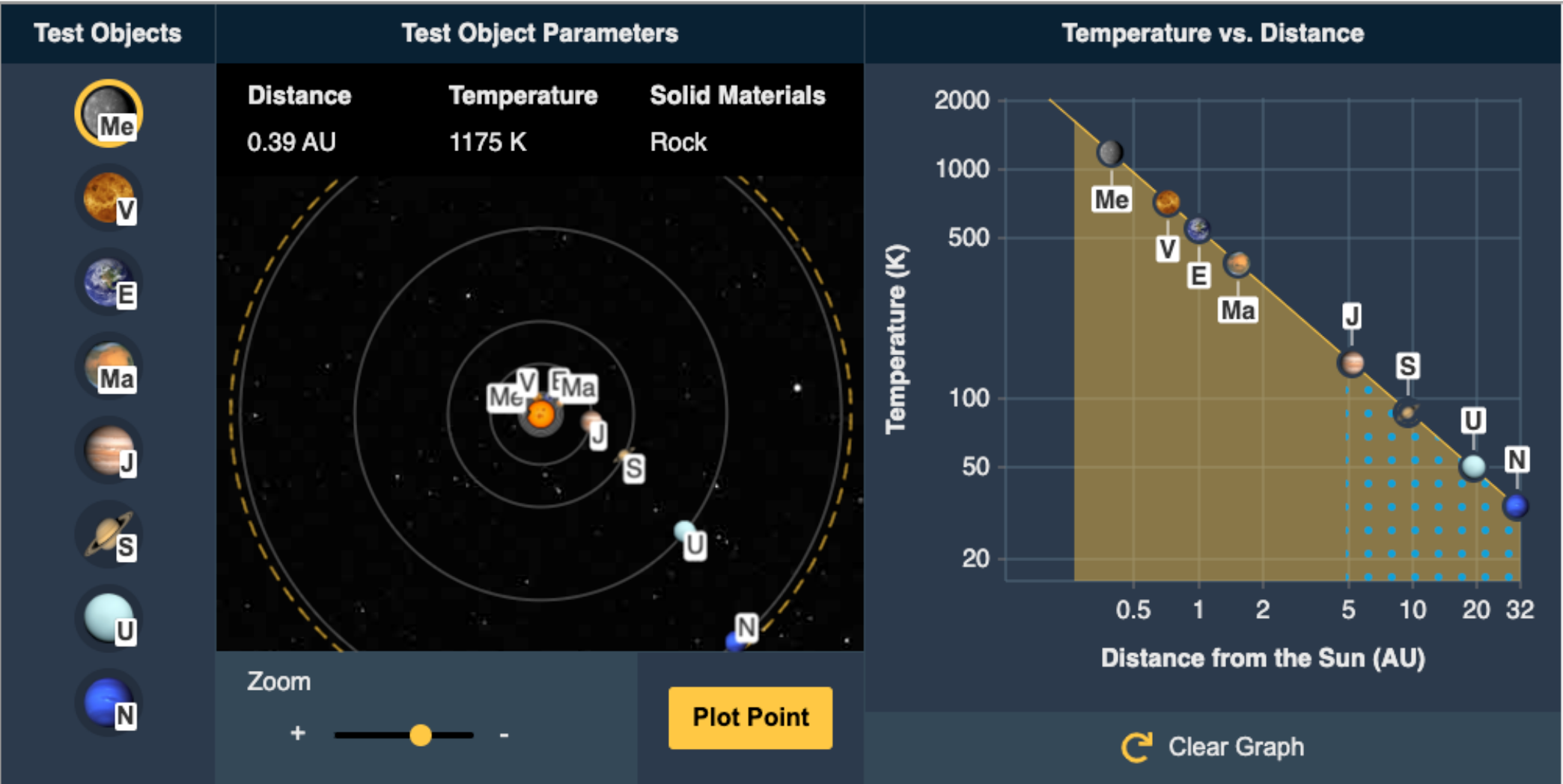}\vspace{-0.5em}
   \caption{}
   \label{fig:planets}
\end{subfigure}
\caption[]{\textbf{(A)} PFOLT simulation screenshot illustrating how students place different materials such as rocks and \textbf{(B)} hydrogen compounds at different locations in the Solar System. When students click ``Plot Point'', a circle or ``X'' appears on the plot indicating whether the material does or does not condense at that location. After plotting enough points, the area below the graph automatically fills in to show the region (and temperature range) over which each material condenses. Yellow corresponds to the region where rocks and metals condense, and the region where hydrogen compounds condense to form ice is overlaid by a blue dot pattern. \textbf{(C)} PFOLT simulation screenshot illustrating how students can place the planets in our Solar System at their current locations. As students drag each planet to its correct location, the planet will appear atop the plots generated in (A) and (B). This allows students to visualize why the terrestrial planets and gas/ice giants have different compositions. Panels (A), (B), and (C) can be compared to the static version from the PFLT presented in Figure \ref{fig:pflt}}
\label{fig:pfolt}
\end{figure}

\begin{figure*}[t!]  
\centering
\includegraphics[width=0.65\textwidth]{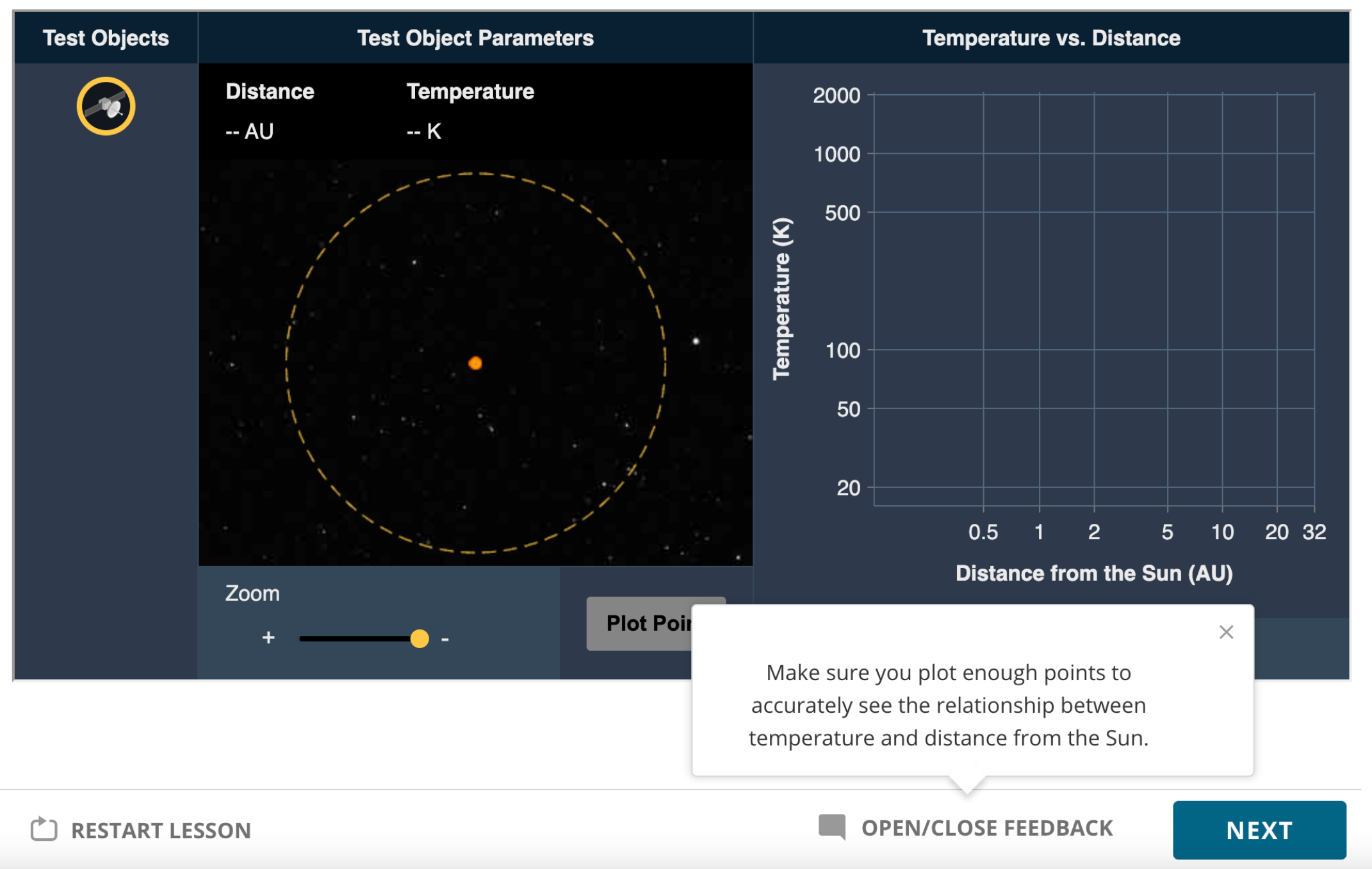}
\caption{PFOLT simulation screenshot showing feedback that appears after students input an incorrect response. In the case shown above, this feedback appears if the student has not plotted enough points to determine the relationship between temperature in the protoplanetary disk and distance from the Sun.
}\label{fig:feedback}
\end{figure*}

\section{Translation of the PFLT to the Planet Formation Online Lecture-Tutorial (PFOLT)}\label{sec:pfolt}

The PFOLT is analogous to the PFLT in its learning outcomes, concepts covered, and question sequence. The PFOLT differs from the pencil-and-paper PFLT in the tasks being asked of students at various places in the activity. For example, in the PFOLT, students discover where different materials (e.g.\ rocks/metals, hydrogen compounds, hydrogen/helium gas) condense in the Solar System by using an interactive simulation to place each material at different distances from the Sun and generating a plot to visualize whether or not each material condenses at that specific location (Figures \ref{fig:rock} and \ref{fig:water}). Students are then taken to a simulation where they can drag each planet in our Solar System to its specific location to better understand which solid materials are available at each location (Figure \ref{fig:planets}). This introduces a level of interactivity the pencil-and-paper PFLT cannot afford, as the PFLT asks students to make interpretations about the relationship between condensation temperature and distance from the Sun using a plot that has already been generated for them (Figure \ref{fig:pflt}). 

Due to the slightly more interactive nature of the PFOLT, the online LT takes students approximately 40 minutes to complete. The PFOLT is also intended to supplement a short lecture on the topic of planet formation, and it can be completed by students either collaboratively or independently as long as they have access to the internet. The PFOLT was developed over the Summer and Fall of 2021 in collaboration with Arizona State University’s (ASU) Center for Education Through eXploration (ETX Center).  

Learning designers at the ETX center have experience outfitting online curriculum with adaptive learning technology. Adaptive learning designs use predetermined rules to provide a learning experience that is tailored to each student’s specific sequence of choices and responses. Prior research has shown this approach to learning design to be very effective, rivaling even human tutoring \citep{vanlehn_relative_2011,kulik_effectiveness_2016}. The key to this effectiveness, as demonstrated by \citet{vanlehn_relative_2011}, is a system that provides feedback to students within the problem solving process, not merely at the end. As students progress through the PFOLT, they receive feedback intended to help them reason through challenges until they reach the correct response, as shown in Figure \ref{fig:feedback}. This adaptive feedback allows students in fully asynchronous (i.e.\ those where instruction is provided solely through pre-recorded material) courses to work through the PFOLT independently, as these students do not have the ability to seek help from their peers or course instructors as they progress through the activity. A complete version of the PFOLT can be accessed free of charge through the NASA Infiniscope website (\url{https://infiniscope.org/}) with the lesson title ``Solar System Formation".

\section{Methods}\label{sec:methods}

\begin{table*}[b!]
\caption{Testing Institution Information}\label{tab:participants}
\begin{tabularx}{\linewidth}{l c l l l}
\toprule
Institution & \# of & Institution Type & Course Modality & Activity \\
 & Courses & & & Implemented \\
\midrule
\multicolumn{5}{c}{\textit{Spring 2022}}\\
\midrule
University of Alabama at & 1 & Public University & Online Asynchronous & PFOLT \\
Birmingham & & & & \\
Glendale Community College$^1$ & 1 & Public Community College & Online Asynchronous & PFOLT \\
Califonia Polytechnic State & 1 & Public University & Online Asynchronous & PFOLT \\
University, San Luis Obispo & & & & \\
University of Colorado Boulder & 1 & Public University & In-Person & PFLT \\
University of Arizona & 1 & Public University & In-Person & PFLT \\
University of Alaska & 1 & Public University & In-Person & PFLT \\
\midrule
\multicolumn{5}{c}{\textit{Fall 2022}}\\
\midrule
University of Michigan & 1 & Public University & Online Asynchronous & PFOLT \\
New Mexico State University & 1 & Public University & In-Person & PFOLT \\
Albion College & 1 & Private Liberal Arts College & In-Person & PFOLT \\
Arizona State University & 2 & Public University & 1 Online Asynchronous, 1 In-Person & PFOLT \\
\bottomrule
\end{tabularx}
\begin{tablenotes}
\item $^1$ Located in Glendale, Arizona
\end{tablenotes}
\end{table*}

\subsection{Settings \& Participants}\label{subsec:settings}

We implemented either the PFLT or PFOLT with students enrolled in eleven different astronomy courses at ten institutions of higher education, between January 2022--December 2022.
Instructors (and their corresponding institutions) were recruited for this study via email correspondence through an astronomy education listserv called 'astrolrner.' The listserv is hosted by the Center for Astronomy Education based out of Steward Observatory at the University of Arizona, but anyone who teaches astronomy or is interested in astronomy education research is able to subscribe. Due to the relatively small nature of the astronomy education community, several of the participating instructors were known to the authors personally, but that was not a requirement for the study. The distribution of institutions in terms of geographic location and institution-type (e.g.\ private versus public) was random, as any instructor who indicated their intent to participate in the study through the listserv was selected.

The implementation sites included one community college and nine four-year colleges and universities with varying degrees of research emphasis. The course modalities were split between in-person and online asynchronous. Originally, we collected data from two additional community colleges with online synchronous and hybrid courses, but they were excluded from the final data set due to low numbers of participants in each category. A complete list of participating institutions from the final data set is provided in Table \ref{tab:participants}.

The students enrolled in ten of the aforementioned courses were predominantly undergraduate non-science majors in the first two years of their undergraduate tenure. The eleventh course in the final data set was an introductory level earth and space science course that enrolled $\sim$70\% (predominantly students in their first year of university) science majors, and $\sim$30\% non-majors. Enrollments in these courses ranged from 8 to upwards of 200 students. This study was approved by Arizona State University’s institutional review board and classified as ``exempt,'' meaning the project did not pose any harm to the study participants and was not subject to further review unless there were significant changes made to the study protocol\footnote{Planet Formation Activity Study, Arizona State University (IRB of Record) ID: STUDY00014402}.

\subsection{Assessments}\label{subsec:assessements}

To evaluate the impact of the PFLT/PFOLT on student learning, participants were given the Planet Formation Concept Inventory (PFCI), a previously validated assessment developed by \citet{simon_development_2019}. A concept inventory is a multiple-choice style instrument that addresses a single topic or closely related set of topics and is written in a way that minimizes scientific jargon and maximizes students’ natural language. Concept inventories differ from traditional multiple-choice assessments in that they use research-based preconceptions as the basis for the incorrect answer choices \citep{Bailey_2009}. We removed 5 questions from the full PFCI that did not cover content presented in either the PFLT or PFOLT. It is important to note that none of the questions from the PFCI were removed in the original analysis conducted by \citet{simon_development_2019}, of which we compare one course’s lecture-only learning gains to those from our study. Because of our item removals from the PFCI, we calculated a Cronbach's alpha on the shortened assessment to verify that it retained satisfactory reliability. Cronbach's alpha is defined as:
\begin{equation}
    \alpha = \frac{K}{K-1} \left( 1 - \frac{\Sigma \sigma_{i}^{2}}{\sigma_x^2} \right)
\end{equation}
where $K$ is the number of test items, $\sigma_i^2$ is the variance of each individual item, and $\sigma_x^2$ is the variance of the full test \citep{Bardar_2006}, with $\alpha \geq 0.70$ considered an acceptable reliability coefficient \citep{Nunnally_1978}.
Using the post-test data from all courses, alpha was 0.736. This is comparable to the original instrument reliability \citep[][Section 3.4]{simon_development_2019}.

The abbreviated PFCI was administered as a pre/post assessment online via QuestionPro. Students completed the PFCI pre-test within the first two weeks of their ASTRO 101 in order to assess their knowledge of planet formation before the instructor covered any related material. They took the PFCI again within a few days of completing either the PFLT or PFOLT. We removed any course section from the final data set where fewer than 50\% of students who took the pre-test were represented in the post-test data. Additionally, to determine how instructors implemented either the PFLT or PFOLT in their respective courses, we developed an instructor implementation survey which was administered via Google Forms at the conclusion of each course. This survey included a series of questions regarding course modality, activity implementation, and the use of other active learning strategies. Responses to the instructor implementation survey informed several of the topics discussed in Section \ref{sec:discuss} as well as the information provided in column 4 of Table \ref{tab:participants}. The survey questions can be found in Appendix A.

\subsection{Normalized gain scores}\label{subsec:lg_meth}

The data reported throughout the remainder of the paper are student responses to the abbreviated PFCI before and after completion of the respective learning activities. Before computing potential learning gains, we removed any students from our sample who completed the PFCI in less than two minutes to avoid the data being skewed by students who did not seriously attempt to answer the questions. To avoid early question bias, we also removed any students who did not answer the last three questions of the PFCI. We also matched students via unique identifiers to ensure that the final data set only included students who took the pre-test, completed either the PFLT or PFOLT, and then took the post-test, hereafter referred to as matched pairs.  It also allowed us to more directly compare any potential learning gains that resulted from completion of the PFLT or PFOLT to those derived from a lecture-only comparison course presented in \citet[][Tables 3 and 4]{simon_development_2019} where learning gains were calculated exclusively with matched pairs data.

Following the procedure outlined in \citet{simon_development_2019}, we computed normalized gain scores for each of the students in the matched pairs data set using the formula:
\begin{equation}
    g_{student} = \frac{post\% - pre\%}{100-pre\%}.
\end{equation}
Additionally, we calculated the average normalized gain score for each of the eleven ASTRO 101 classes in our sample:
\begin{equation}
    g_{class} = \frac{<\!M\!>post\%\ - <\!M\!>pre\%}{100\ - <\!M\!>pre\%}
\end{equation}
where <M> is the mean pre-/post-test score. We used student-level (rather than course-level) gain calculations when employing an analysis of variance as described in Section \ref{subsec:stats} to look for statistical significance between modalities and between the activity types. Because $g$ is undefined for students with perfect pre-test scores, these students were excluded and our number of students differ very slightly depending on whether we are using $g_{class}$ or $g_{student}$ for our analysis.

\subsection{Inferential statistics}\label{subsec:stats}

There are two key questions in our analysis. First, are student learning gains following the PFOLT comparable to those from the in-person implementation of the pencil-and-paper PFLT? Second, do student learning gains from classes that used the PFOLT exceed those of traditional lecture classes, which used no additional active learning activities? To answer these questions, we employed a one-way between group analysis of variance (ANOVA) with post hoc testing using Tukey’s HSD (honestly significant difference) to identify statistically significant paired comparisons \citep{Toothaker_1993}. Because the PFOLT was tested in both in-person and asynchronous modalities (see Table \ref{tab:participants}), we treat these two course instructional modalities as distinctly different in our analysis. Finally, our lecture comparison data come from a previously published study \citep[][Table 3]{simon_development_2019}. Thus, we only had access to course-level summary data (i.e.\ mean, standard deviation, and number of students). An ANOVA requires student-level data. Therefore, we simulated student-level data with these characteristics using the \texttt{rnorm()} function in R before performing the ANOVA.

\begin{figure}[tb!]  
\centering
\includegraphics[width=\columnwidth]{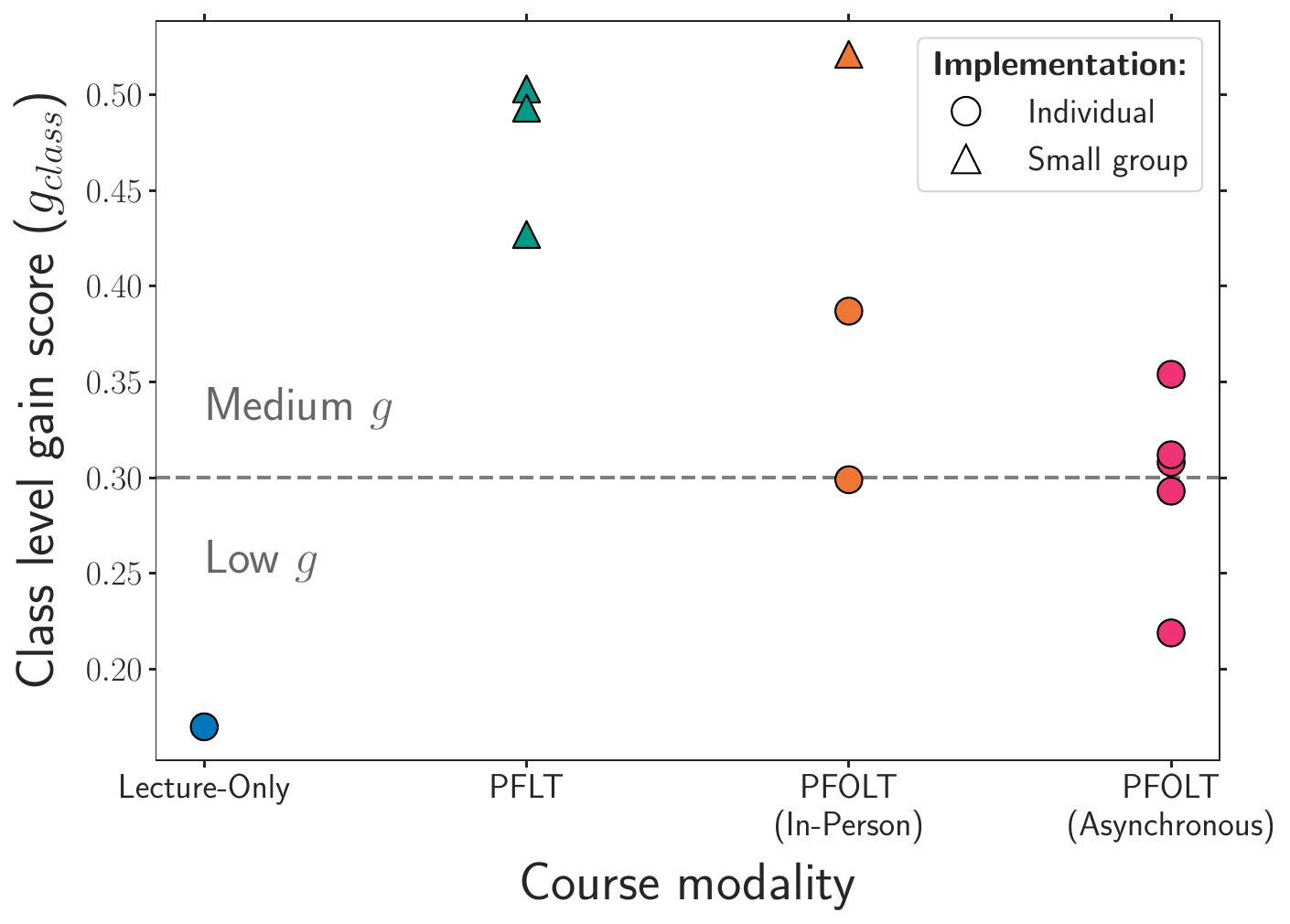}
\caption{Class level gain score ($g_{class}$) for each instructional method. $g_{class}$ for the lecture-only condition was taken from \citet[][Table 3]{simon_development_2019}. The lecture-only course and the courses that utilized the traditional PFLT were all taught in-person. PFOLT $g_{class}$ results are split into two categories: in-person and asynchronous to denote online asynchronous. Whether the intervention was conducted individually or in small groups is marked by either circles or triangles (the lecture-only course is also marked with a circle). The gray, dashed line separates low normalized gain ($g_{class} < 0.3$) from medium normalized gain ($0.3 < g_{class} < 0.7$).
}\label{fig:plot}
\end{figure}

\begin{table*}[tb!]
\caption{Matched-pairs measured learning gains}\label{tab:matchedpairs}
\begin{tabularx}{\linewidth}{l l l c c c c c c}
\toprule
Institution & Modality & Activity & \# of & Pre-test & Pre-test & Post-test & Post-test & $g_{class}$ \\
 & & Implemented & Matched & <M> & <SD> & <M> & <SD> & \\
 & & & Pairs$^1$ & (\%) & (\%) & (\%) & (\%) & \\
\midrule
University of Alaska & In-Person & PFLT & 15 & 63.6 & 15.9 & 79.1 & 18.4 & 0.427 \\
University of Colorado Boulder & In-Person & PFLT & 84 & 49.8 & 18.4 & 75.1 & 18.9 & 0.503 \\
University of Arizona 1 & In-Person & PFLT & 198 & 48.8 & 17.5 & 74.0 & 17.5 & 0.493 \\
University of Alabama at & Asynchronous & PFOLT & 34 & 45.9 & 16.6 & 62.5 & 23.1 & 0.308 \\
Birmingham & & & & & & & & \\
California Polytechnic State & Asynchronous & PFOLT & 64 & 48.5 & 16.7 & 66.8 & 17.7 & 0.354 \\
University, San Luis Obisbo & & & & & & & & \\
Glendale Community College$^2$ & Asynchronous & PFOLT & 12 & 48.9 & 14.7 & 63.9 & 25.7 & 0.293 \\
New Mexico State University & In-Person & PFOLT & 14 & 53.8 & 13.2 & 67.6 & 20.0 & 0.299 \\
Albion College & In-Person & PFOLT & 14 & 41.0 & 15.7 & 63.8 & 16.5 & 0.387 \\
University of Michigan & Asynchronous & PFOLT & 21 & 49.2 & 20.0 & 65.1 & 23.8 & 0.312 \\
Arizona State University 1 & In-Person & PFOLT & 149 & 58.7 & 19.1 & 80.2 & 16.9 & 0.521 \\
Arizona State University 2 & Asynchronous & PFOLT & 150 & 57.2 & 17.1 & 66.6 & 20.7 & 0.219 \\
University of Arizona 2$^3$ & In-Person & Lecture-Only & 40 & 51.0 & 14.9 & 60.1 & 20.2 & 0.170 \\
\bottomrule
\end{tabularx}
\begin{tablenotes}
\item $^1$ Number of students where we were able to match pre and post-tests
\item $^2$ Located in Glendale, Arizona
\item $^3$ Data based on \citet{simon_development_2019}
\end{tablenotes}
\end{table*}

\section{Results}\label{sec:results}

\subsection{Learning gains}\label{subsec:lg_res}

Table \ref{tab:matchedpairs} provides a summary of learning gains data by class. This includes mean <M> and standard deviation <SD> values for pre- and post-tests and the normalized gain score for each class, $g_{class}$. The table also includes course modality, activity implemented, and number of matched pairs. In Figure \ref{fig:plot} we show $g_{class}$ for the different activity types (PFLT/PFOLT) while also highlighting implementation strategy (whether students completed the respective activity independently or in small groups). In both Table \ref{tab:matchedpairs} and Figure \ref{fig:plot}, we include the lecture-only learning gains reported in \citet{simon_development_2019}, where students took the full PFCI in an ASTRO 101 course with no active learning interventions. Note that learning gains in this lecture-only condition were measured using the full PFCI, in contrast to the other classes which were measured using a 15-question  subset of the same assessment. Three categories of normalized gain scores are defined by \citet{hake_interactive-engagement_1998} and \citet{prather_teaching_2009}: low ($g < 0.3$), medium ($0.3 < g < 0.7$), and high ($g > 0.7$). A dashed line in Figure \ref{fig:plot} denotes the separation between low and medium normalized gain scores.

\begin{table*}[hbt!]
\caption{Descriptive condition-level learning gain statistics$^1$}\label{tab:stats}
\begin{tabularx}{\linewidth}{c l c c c c c c c}
\toprule
 & Activity Implemented & & \# of Matched Pairs$^2$ & $g_{student}$ <M> & $g_{student}$ <SD> & $g_{student}$  Min & $g_{student}$  Max & $g_{student}$ <SE$^3$> \\
\midrule
 & PFLT & & 297 & 0.480 & 0.360 & -1.750 & 1.000 & 0.021 \\
 & PFOLT (In-Person) & & 175 & 0.506 & 0.350 & -1.000 & 1.000 & 0.026 \\
 & PFOLT (Asynchronous) & & 280 & 0.256 & 0.448 & -2.000 & 1.000 & 0.027 \\
 & Lecture-Only$^4$ & & 40 & 0.170 & 0.245 & -0.362 & 0.577 & 0.039 \\
\bottomrule
\end{tabularx}
\begin{tablenotes}
\item $^1$ Values shown are averages of student-level data
\item $^2$ Number of students where we were able to match pre and post-tests
\item $^3$ Standard error
\item $^4$ Data based on \citet{simon_development_2019}
\end{tablenotes}
\end{table*}

\begin{table}[b!]
\caption{One-way ANOVA summary}\label{tab:anovaSum}
\begin{tabularx}{\linewidth}{l S c c c c}
\toprule
& {Sum of}& df$^1$ & Mean& F$^2$ & Significance \\
& {Squares} & & Square & & \\
\midrule
Between& 11.734 & 3 & 3.911 & 26.11 & < .001 \\
Groups & & & & & \\[0.3em]
Within & 118.059 & 788 & 0.150 & & \\
Groups & & & & & \\[0.3em]
Total & 129.793 & 791 & & &\\
\bottomrule
\end{tabularx}
\begin{tablenotes}
\item $^1$ Degrees of freedom
\item $^2$ F statistic
\end{tablenotes}
\end{table}

\begin{table*}[t!]
\caption{Pairwise comparisons}\label{tab:pairwise}
\begin{tabularx}{\linewidth}{c l l S S S S}
\toprule
 & \multicolumn{2}{c}{Activity Implemented} & & \multicolumn{2}{c}{95\% Confidence Interval} & \\
\cmidrule{2-3} \cmidrule{5-6}
 & {Condition 1} & {Condition 2} & {Mean Difference} & {Lower Bound} & {Upper Bound} & {Significance} \\
 & & & {(Condition 1 - Condition 2)} & & & \\
\midrule
 & PFLT & PFOLT (In-Person) & -0.026 & 0.069 & -0.121 & 0.892 \\
 & PFLT & PFOLT (Asynchronous) & 0.224 & 0.307 & 0.141 & <0.001 \\
 & PFLT & Lecture-Only & 0.310 & 0.477 & 0.142 & <0.001 \\
 & PFOLT (In-Person) & PFOLT (Asynchronous) & 0.250 & 0.346 & 0.154 & <0.001 \\
 & PFOLT (In-Person) & Lecture-Only & 0.336 & 0.511 & 0.161 & <0.001 \\
 & PFOLT (Asynchronous) & Lecture-Only & 0.086 & 0.254 & -0.083 & 0.558 \\
\bottomrule
\end{tabularx}
\end{table*}

\subsection{ANOVA results}\label{subsec:anova}

A comparison of student-level learning gains across instructional methods was significant overall with the F statistic, F(3, 788) = 26.11; $p < .001$. Speaking to our first question of interest (how does student learning compare when using the PFLOT versus using the PFLT), post hoc testing shows the PFLT to have significantly higher learning gains ($p < .001$) than the asynchronous implementation of the PFOLT, but there is virtually no difference ($p = .89$) between the PFLT and the in-person implementation of the PFOLT. Regarding our second question (how does the PFOLT compare to lecture-only instruction), post hoc testing shows that the in-person PFOLT had significantly higher learning gains ($p < .001$) than lecture. However, the asynchronous PFOLT was not significantly different from lecture ($p = .56$). Finally, testing also indicates that the in-person implementation of the PFOLT was significantly more effective than the asynchronous implementation ($p < .001$). The complete ANOVA results and relevant descriptive statistics are presented in Tables \ref{tab:stats}-\ref{tab:pairwise}.

\section{Discussion}\label{sec:discuss}

\subsection{Exploration of learning gains}\label{subsec:lg_dis}

When the PFOLT was implemented in-person, the learning gains were comparable to the PFLT. However, when implemented asynchronously, the learning gains were comparable to the lecture-only group. One likely explanation for this pattern can be attributed to the value of small group learning. Studies find that students who work in small groups showed significantly greater gains on conceptual questions than students who worked individually \citep{gokhale_collaborative_1995,adams_learning_2002}. When working together in small groups, ASTRO 101 students, who are often at varying levels of discipline knowledge and ability, are better able to reason through a problem when presented with other perspectives or interpretations of their peers. The higher learning gains among courses implementing the PFLT/PFOLT in small groups further highlights the importance of collaborative learning. In addition to underscoring one of the fundamental pedagogical tenets of LTs, this finding is also consistent with findings from other research in active learning, particularly the ICAP (Interactive, Constructive, Active, and Passive) framework \citep{chi_icap_2014} which found ``interactive'' learning, i.e.\ co-construction of knowledge, to be the most effective form of active learning.

In the case of the PFOLT, however, the single course that implemented the activity both in small groups and in-person was the only course in the final data set that enrolled predominantly science majors (see Section \ref{subsec:settings}). Despite being a majors-dominant course, the course was still at the introductory level and did not have any science prerequisite. It is expected that science majors will out-perform non-majors and, indeed, \citet{simon_development_2019} found that science majors’ normalized gain scores were significantly higher than those of non-science majors on the PFCI. Had the course been made up of entirely non-science majors, we predict the gain scores would be lower to some extent. To this end, we likely cannot attribute this class’ high gain score ($g_{class} = 0.521$) to small group collaboration alone. 

Additionally, prior research indicates that the quality of instructor implementation can be the most crucial factor in determining gain scores \citep{prather_teaching_2009,wallace_study_2012}. This factor impacted our results in two distinct ways. First, the highest learning gains were found in classes where instructors facilitated small group collaboration (Figure \ref{fig:plot}). Second, the presence and quality of prior instruction likely played a role in the lower-than-expected learning gains for the asynchronous PFOLT courses. Lecture-tutorials, conventionally, supplement lecture instruction on a given topic. In contrast to the PFLT classes and most in-person PFOLT classes, not all of the asynchronous classes included instruction on planet formation prior to the PFOLT. Of the five asynchronous courses in our final data set, one preceded the PFOLT with a separate interactive digital tutorial. The other four had lower quality prior instruction, one providing no prior instruction at all, and the other three providing asynchronous videos or readings that were recommended but not required. Although not all of these videos and readings were trackable, from the data that were available, less than half of students viewed these materials, thus beginning the PFOLT without any pre-activity engagement. 

Both of these variations in quality of implementation complicate the interpretation of our results. For example, since all of the PFLT data come from in-person classes that implemented that activity after a lecture on the topic, and with small groups, and, conversely, no fully online class employed small group instruction (e.g.\ through webinar break out groups) we cannot fully disentangle implementation and activity-type. Similarly, because the measured learning gains reflect gains from \textit{both} the lecture and the LT, in classes without any required prior instruction, the LT itself is responsible for relatively greater learning (i.e.\ some portion of what would otherwise have been learned in the lecture portion). Not to mention the value found from repetition and reinforcement of concepts when a LT is preceded by a lecture or other instruction.

In summary, while this study does not find clear, statistically significant differences between lecture only and the PFOLT in all implementations, the higher gain scores observed for the online asynchronous condition when compared to traditional lecture (despite limited pre-instruction) indicate that the PFOLT is worthy of being used as a tool to teach planet formation in ASTRO 101 courses online. Finally, our results also underscore the value of small group learning and highlight a recurring challenge in asynchronous online learning settings to find ways to build in opportunities for peer-learning.

\subsection{Activity improvements}\label{subsec:limits}

In our instructor implementation survey, we requested feedback on the implementation of either the PFLT or PFOLT in their classes. Two instructors suggested that the redundancy of plotting points on the graphs in the PFOLT caused students to lose interest. In future versions of the PFOLT, we will program the activity such that the graphs automatically fill in earlier than they currently do, immediately after students demonstrate an understanding of the relationship they are intended to plot.

Even though we typically observed higher PFOLT learning gains in in-person courses, we anticipate the PFOLT will be used predominantly in ASTRO 101 courses online. Since the PFOLT is designed to be used in asynchronous courses where students often work independently, outfitting the activity with a more complex, intelligent tutoring system (ITS) would likely lead to more profound student learning than what we currently observe. Unlike human tutoring, computer-based tutoring is traditionally separated into two technological types: answer-based and step-based
\citep{vanlehn_relative_2011}. As it stands, the PFOLT falls under the answer-based category, which gives students immediate feedback and hints based on their answer choices. Adding a step-based ITS would provide students with feedback and hints along each step of the problem-solving process, similar to conversing directly with a peer. Alternatively, the inconsistency of prior instruction could be addressed by building in a standardized pre-recorded lecture.

Furthermore, a meta-analysis \citep{wisniewski_power_2020} of more than 400 research studies looking at the effects of feedback on student learning found that the ``cognitive complexity'' of adaptive feedback directly relates to the effectiveness of the feedback. The three categories of complexity ranging from least to most complex are: task level feedback, process level feedback, and self-regulation feedback. Currently, the PFOLT utilizes task-level feedback, providing students  with responses regarding whether a task was done correctly rather than presenting students with suggestions and strategies concerning how to complete each task. We plan to work with the ETX center at ASU to integrate an ITS into the PFOLT specifically designed to offer more process and/or self-regulation based feedback with the goal of further increasing student learning.

\section{Summary \& Conclusions}\label{sec:summary}

An uptick in online course enrollments coupled with the COVID-19 pandemic put a spotlight on the need for additional effective, research-based, curricular materials that lead to more lasting conceptual change. As one contribution toward this overarching objective, we developed and explored the efficacy of a novel, digital LT intended to teach planet formation in online ASTRO 101 courses. We utilized the previously validated PFCI to compare student and course-level learning gains between lecture-only courses, courses that implemented the PFOLT, and those that implemented the traditional pencil-and-paper PFLT. Several previous efforts conclusively demonstrate that LTs are incredibly effective at increasing student learning on a myriad of topics when compared to lecture alone. To date, however, all available LTs for ASTRO 101 are pencil-and-paper based, having been developed exclusively for courses taught in-person. 

Overall, our results show learning gains from these pencil-and-paper LT (PFLT) to be statistically indistinguishable from the in-person implementation of the PFOLT and show that both of these conditions exceed gains from lecture-only instruction. However, when implemented asynchronously, learning gains from the PFOLT were lower and not statistically distinct from the lecture-only condition. These results are qualified by important differences in instructor implementation, including learning in small groups versus individual work and the presence and quality of pre-LT instruction. The highest learning gains for the PFOLT were also found in an introductory course primarily intended for science majors, whereas all other data came from courses for non-science majors. While improvements can be made to improve the online LT in the future, the current version still outperforms traditional lecture (for in-person, small group implementations), and can be used as a tool to teach planet formation effectively. 

In a future research study, we plan to revisit the question of whether LTs can be effective in online, asynchronous classes. This work will be done following revisions to the PFOLT and with tighter controls on pre-activity instruction. We will update the PFOLT as described in Section \ref{subsec:limits}, including addressing plotting redundancies and improving automatic feedback. To better ensure similar pre-activity instruction across testing sites, we will embed a pre-recorded lecture video that will precede the interactive component of the activity. The potential benefits from LT-style instruction in asynchronous online classes are compelling, but the inherent differences in that modality raise real concerns about how to effectively translate a proven in-person active learning strategy.

\section{Declarations}
Data from this study can be obtained from the authors upon reasonable request.

\subsection{List of abbreviations}
\begin{itemize}
    \item ANOVA: Analysis of Variance
    \item ASU: Arizona State University
    \item ETX: Education Through eXploration
    \item $g$: Normalized Learning Gain
    \item HSD: Honestly significant difference
    \item ICAP: Interactive, Constructive, Active, and Passive
    \item ITS: Intelligent Tutoring System
    \item LT: Lecture-Tutorial
    \item M: Mean
    \item PDR: Pedagogical Discipline Representation
    \item PFCI: Planet Formation Concept Inventory
    \item PFLT: Planet Formation Lecture-Tutorial
    \item PFOLT: Planet Formation Online Lecture-Tutorial
    \item SD: Standard Deviation
    \item SE: Standard Error
\end{itemize}

\subsection{Ethical Approval (optional)}
This study was approved by Arizona State University’s institutional review board and classified as ``exempt'' under IRB of Record ID: STUDY00014402.

\subsection{Consent for publication}

Not applicable.

\subsection{Competing Interests}

The author(s) declare that they have no competing interests.

\subsection{Funding}

This material is based upon work supported by the NASA SMD Exploration Connection under Award Number: NNX16AD79A.

\subsection{Author's Contributions}
\noindent \textit{H.\ N.\ Archer}: Data curation, formal analysis, investigation, methodology, resources, software, validation, visualization, writing -- original draft.\\ \textit{M.\ N.\ Simon}: Conceptualization, data curation, funding acquisition, investigation, methodology, project administration, resources, supervision, validation, writing -- original draft.\\ \textit{C.\ Mead}: Conceptualization, methodology, formal analysis, software, validation, visualization, writing -- original draft.\\ \textit{E.\ E.\ Prather}: Resources, writing -- review and editing.\\ \textit{M.\ Brunkhorst}: Data curation, resources, software, visualization.\\ \textit{D.\ Hunsley}: Data curation, resources, software, visualization.

\section{Acknowledgements}

The authors would like to acknowledge each of the instructors who allowed us to implement either the PFLT or PFOLT in their courses, along with the students who contributed their efforts to this work.

\bibliography{paper-refs}

\section{Appendix A}\label{sec:appendix}

\subsection{Instructor Implementation Survey}

\textbf{Instructor Information}
\vspace{-1em}
\begin{enumerate}
    \item Instructor Name
    \item Instructor Email
    \item Instructor Institution
    \item How many students were enrolled in your course(s)?
    \vspace{-1em}
    \begin{itemize}
        \item[$\circ$] <25
        \item[$\circ$] 25-50
        \item[$\circ$] 50-100
        \item[$\circ$] >100
    \end{itemize}
    \vspace{-1em}
    \item What was the course modality?
    \vspace{-1em}
    \begin{itemize}
        \item[$\circ$] Asynchronous online
        \item[$\circ$] Synchronous online
        \item[$\circ$] In-person
        \item[$\circ$] Hybrid
        \item[$\circ$] Other [please explain]
    \end{itemize}
    \vspace{-1em}
    \item Which activity did you implement?
    \vspace{-1em}
    \begin{itemize}
        \item[$\circ$] Online lecture-tutorial
        \item[$\circ$] Pencil-and-paper lecture-tutorial
        \item[$\circ$] Both
    \end{itemize}
\end{enumerate}
\vspace{-1em}
\noindent \textbf{Pencil-and-paper lecture-tutorial}
\vspace{-1em}
\begin{enumerate}
    \item Please describe how you implemented the lecture-tutorial (e.g.\ student groups, students working individually, students working in zoom breakout rooms). Did you implement it all at once? Break it into sections?
    \item Please provide feedback regarding how the lecture-tutorial could be improved (or what you liked about it).
\end{enumerate}
\vspace{-1em}
\noindent \textbf{Online lecture-tutorial}
\vspace{-1em}
\begin{enumerate}
    \item Please describe how you implemented the online lesson in your course.
    \item Would you use this online lesson again in your course?
    \vspace{-1em}
    \begin{itemize}
        \item[$\circ$] Yes
        \item[$\circ$] No
    \end{itemize}
    \vspace{-1em}
    \item Please provide feedback regarding how the online lecture-tutorial could be improved (or what you liked about it).
\end{enumerate}
\vspace{-1em}
\noindent \textbf{General Planet Formation Teaching Questions}
\vspace{-1em}
\begin{enumerate}
    \item In your course, do you implement any active learning activities beyond the in-person tutorial or online lesson (e.g.\ think-pair-share questions, additional tutorials)?
    \vspace{-1em}
    \begin{itemize}
        \item[$\circ$] Yes
        \item[$\circ$] No, lecture-only
    \end{itemize}
    \vspace{-1em}
    \item If yes, please briefly explain.
    \item When teaching planet formation, did you implement any active learning strategies beyond the pencil-and-paper/online lecture-tutorial?
    \vspace{-1em}
    \begin{itemize}
        \item[$\circ$] Yes
        \item[$\circ$] No, lecture-only
        \item[$\circ$] I did not cover planet formation beyond what was in the lecture-tutorial
    \end{itemize}
    \vspace{-1em}
    \item If yes, please briefly explain.
    \item By the time my students have taken the post-test, they have learned the following concepts in my class: SELECT ALL THAT APPLY.
    \vspace{-1em}
    \begin{itemize}
        \item[$\square$] The definition of an exoplanet
        \item[$\square$] The definition of a solar system
        \item[$\square$] The definition of a star
        \item[$\square$] The definition of a planet
        \item[$\square$] The definition of a dwarf planet
        \item[$\square$] Planetary motion/orbits
        \item[$\square$] The nebular theory
        \item[$\square$] Accretion: planetesimals into planets
        \item[$\square$] The composition of the rocky planets and gas giant planets
        \item[$\square$] Condensation temperature and/or the snowline
        \item[$\square$] Basic concept of planetary migration
        \item[$\square$] The formation of the Universe -- the Big Bang
        \item[$\square$] The size and scale of the Universe (e.g.\ what is a galaxy versus a solar system)
    \end{itemize}
\end{enumerate}

\end{document}